\documentclass[prb,twocolumn,showpacs,superscriptaddress]{revtex4}
\usepackage{graphicx,psfrag,amsmath,amssymb,amsfonts,bbm,latexsym,color,dcolumn}

\begin{document}

\newcommand\ket[1]{|#1\rangle}
\newcommand\bra[1]{\langle#1|}
\newcommand\braket[2]{\left\langle#1\left|#2\right.\right\rangle}

\title{Entanglement detection for electrons via witness operators}

\author{Lara Faoro}
\affiliation{Department of Physics and Astronomy, Rutgers University, 136 Frelinghuysen
Rd, Piscataway 08854, New Jersey, USA}
\author{Fabio Taddei}
\affiliation{NEST CNR-INFM \& Scuola Normale Superiore, I-56126 Pisa, Italy}

\date{\today}

\begin{abstract}
We discuss an implementation of the entanglement witness, 
a method to detect entanglement with few local measurements, in systems where 
entangled electrons are generated both in the spin and orbital degrees of freedom.
We address the efficiency of this method in various setups, including two different particle-hole entanglement structures, and we demonstrate that it can also be used to infer information on the possible dephasing afflicting the devices. 
\end{abstract}

\maketitle

\section{Introduction}
Quantum entanglement is a key resource in the theory of quantum information, 
including quantum cryptography \cite{Ekert91}, quantum communication \cite{Bennett-tel} and quantum computation \cite{Steane98,NielsenBook}. 
In the last few years significant efforts have been undertaken 
in order to engineer devices capable of generating maximally entangled 
states. 
In particular, in view of their integrability and scalability,
considerable attention has been devoted to the 
generation of entanglement for electrons in solid state systems (for a review see Refs.~\onlinecite{Beenakker05-2,Burkard05}).
Proposals aim at producing pairs of entangled 
electrons either in the spin or in the orbital degrees of freedom. 
Some of them are based on the generation of Bell states by means of electron-electron 
interaction \cite{Loss00,Recher01,Lesovik01,Bouchiat03,Samuelsson03,Saraga03}. 
Only recently it has been shown that in a mesoscopic multiterminal 
conductor entanglement can be also produced in the absence of 
electron-electron interaction \cite{Beenakker03,Faoro04,Samuelsson04,Lebedev04,DiLorenzo05,Signal05,Frustaglia05} and
through time-dependent potentials
\cite{Samuelsson05,Beenakker05,Lebedev05}.

Unfortunately, electronic entangled states are fragile. 
Indeed, it is well known that the unwanted interaction with 
the environment might  
lead to loss of quantum coherence and eventually to the destruction of 
the entanglement between the electronic quantum states. 
In view of experimental implementations, it is therefore of crucial importance the search for simple procedures able  
to assess the entanglement {\em actually} generated in a given device (the {\em entangler}) under non vanishing environmental couplings. 

Several strategies have been proposed in order to detect 
electronic entanglement in mesoscopic conductors.
One of these schemes makes use of an additional 
beam splitter, to be attached to the entangler generating pairs 
of spatially separated entangled electrons, which 
reveals the presence of entanglement through a measurement of the correlations 
of the current fluctuations in the exit leads \cite{Burkard00}.
The possibility of evaluating a lower bound for the entanglement of formation was later demonstrated in Ref.~\onlinecite{Burkard2003} and, recently, a generalization in terms of a Hong-Ou-Mandel interferometer was presented \cite{Giovannetti06}.
The influence of Rashba spin-orbit coupling \cite{Egues02,Egues05}, rotating magnetic fields \cite{Zhao05} and dephasing contacts \cite{Sanjose06} were also investigated.
In addition, it has been shown that signatures of 
entanglement in this prototype setup can be detected through the 
full counting statistics \cite{Taddei02}.
A second scheme makes use of
Bell inequality tests, which have been formulated both in terms of current correlations at different terminals \cite{Kawabata01,Samuelsson03,Beenakker03,Bouchiat03,Samuelsson04,
Lebedev04,Lebedev05,Samuelsson05} and by resorting to the full counting statistics \cite{Faoro04,Prada05}.
More recently, the following other schemes have been considered.
For a superconductor-dot entangler, spin current correlations have been shown to give a characterization of the entangler operation \cite{Sauret05}.
A one-to-one relation between the entanglement production rate of electron-hole pairs and the spin-resolved shot noise has been derived in Ref.~\onlinecite{Michaelis06} in the case where the elastic scattering is spin-independent.
A proposal for detecting energy entanglement in a normal-superconducting junction, making use of a Mach-Zehnder interferometer, has been also presented \cite{Bayandin06}.
Furthermore, a quantum state tomography scheme has been put forward \cite{Samuelsson06}.

In this paper we shall show that it is possible to implement the idea of
entanglement {\em witness} operators \cite{footnoteNo1}
to detect entanglement for electrons in 
multiterminal mesoscopic conductors afflicted by ``noise''.
The concept of entanglement witness has been introduced only 
recently in quantum information theory \cite{Lewenstein01} and experimentally applied to optical systems \cite{Barbieri03}. 
It has been shown that, given a class of states 
(containing both separable and entangled states), it is possible to 
construct an Hermitian operator, called 
{\em witness}, having the remarkable property that its expectation 
value is positive for all the separable states contained in 
that class. 
If this is true {\em only} for separable states, the witness is said to be {\em optimal} \cite{Lewenstein00}.
As a consequence, {\em all} entangled states are singled out by 
negative expectation values of the optimal witness operator. 
Although the entanglement witness detection is not 
a measure of entanglement, it represents an 
efficient method to detect entangled states.

In order to implement optimal witness detection for electrons in
mesoscopic structures one needs:  (i) to identify the class of 
states that is generated by the given setup, (ii) to construct the 
optimal witness operator, (iii) to be able to measure its expectation value.
Given a physical setup and a proper phenomenological model for the ``noise'' 
afflicting the device, the class of states 
can be easily identified. Also the construction of the optimal entanglement witness
does not present much difficulties once the separability criterion 
\cite{peres-1996-77,Horodecki96} can be used.
Unfortunately, the measurement of the witness operator is not always an easy task.
Furthermore, the witness operator can be measured only 
when it does not depend on the unknown parameters describing the presence of ``noise''. 
In that case we shall demonstrate, for different setups, that the mean value of the witness 
operator can be determined through the measurement of current cross-correlations.

The effectiveness of the entanglement witness stems from the fact that the set 
of cross-correlated measurements required to determine its expectation 
value is fixed, irrespective of the degree of ``noise'' present, and the number of such measurements is small.
Typically, from 3 to 5 analyzer's settings are necessary, corresponding to a number of current cross-correlations of the order of 10.
In this respect it might be preferable to a Bell inequality test, which, in the presence of ``noise'', demands an unknown and possibly large number of analyzer's settings to be tested in order to search for violations. 
Furthermore, the Bell inequality test is only a sufficient criterion for the detection of entanglement \cite{Terhal02}, while the witness, if optimal, is both sufficient and necessary.
Witness entanglement detection might also be preferable to quantum state 
tomography, especially when one is not interested in acquiring information 
over the {\em whole} density matrix of the system, but simply 
wishes to know if some entanglement has survived at the exit of the setup.

The paper is organized as follows. 
In Sec.~\ref{witnesspar} we introduce 
the concept of witness operator and discuss how the latter can be 
easily constructed by using the separability criterion 
\cite{peres-1996-77,Horodecki96}. 
In Sec.~\ref{ideal} we describe the details of witness entanglement detection
by applying it to a prototype setup where electron spin singlet and 
triplet states are initially generated and subsequently corrupted by 
decoherence due to spin dephasing and spin relaxation in the wires.   
We then discuss under 
which conditions the measurement of the witness operator can be achieved 
and we show that the expectation value of the witness operator can be 
expressed in terms of current cross-correlators. We also explain why 
entanglement witness detection might be more efficient than usual 
Bell inequality tests.
In Sec.~\ref{ph} we then proceed to construct the witness operators for two 
realistic setups generating particle-hole entanglement. 
Specifically, in Sec.~\ref{Hall} we consider  
the quantum Hall bar system proposed in Ref.~\onlinecite{Beenakker03}
while in Sec.~\ref{HBT} we discuss 
the electronic Hanbury-Brown and Twiss interferometer proposed in Ref.~\onlinecite{Samuelsson04}.
Sec.~\ref{conc} is finally devoted to the conclusions and future perspectives.

\section{The witness}
\label{witnesspar}
Let $\rho$ be the state of a bipartite system acting on the Hilbert space 
${\mathcal{H}=\mathcal{H}_{\text{A}}\otimes\mathcal{H}_{\text{B}}}$ and let us denote with A and B the two subsystems. 
The state $\rho$ is called entangled \cite{Werner} (or non separable) if it {\em cannot} be written as a convex combination of product states, i.e. as:
\begin{equation}
\rho=\sum_k p_k |a_k,b_k \rangle \langle a_k,b_k| \;
\label{w1}
\end{equation}
where ${p_k \ge 0}$ and ${|a_k,b_k \rangle \equiv |a_k \rangle_A \otimes |b_k \rangle_B}$ are product vectors. Conversely, the state $\rho$ is called 
separable (or not entangled) if it can be written in the form given in Eq. (\ref{w1}).
 
It is well known that, limited to the case of 
$\mathcal{H}=\mathbb{C}^2\otimes\mathbb{C}^2$ or $\mathcal{H}=\mathbb{C}^2\otimes\mathbb{C}^3$, it is possible to determine whether $\rho$ is entangled or not
by calculating the partial transpose 
(i.e. the transpose with respect to one of the subsystems). In fact, according to the  separability criterion 
\cite{peres-1996-77,Horodecki96},
the state $\rho$ is separable if and only if its partial transpose 
is positive.  
Alternatively, one can distinguish between an entangled state $\rho$ and 
all the separable states by introducing the concept of {\em entanglement witness} 
\cite{Horodecki96}. 
Given an entangled state $\rho$,
an entanglement witness is indeed an Hermitian operators $W$ 
having the remarkably propriety that ${\text{Tr}(W\rho) <0}$ 
and ${\text{Tr}(W\rho_{\text{sep}}) \ge 0}$ for 
all separable states $\rho_{\text{sep}}$. 
The negative expectation value 
of $W$ is hence a signature of entanglement and  
the entangled state $\rho$ is said to be {\em detected} by the witness. 

Given a class of states 
$\rho_{\Lambda}$ (containing both separable and entangled states) we define as optimal \cite{Lewenstein00} the witness operator $W_{\text{opt}}$ that is able 
to detect {\em all} the entangled states present in the class 
$\rho_{\Lambda}$. As a result,
whenever a class of states $\rho_\Lambda$ and the optimal entanglement 
witness for that class are given, it is possible to determine whether a state 
${\rho \in \rho_\Lambda}$ is entangled by simply checking if 
${\langle W_{\text{opt}} \rangle = \text{Tr} (W_{\text{opt}} \rho)}$ is 
negative.

As one can expect, finding an optimal witness for a given class of states 
is, in general, not a trivial task. 
However, the optimal witness construction can 
be greatly simplified by resorting to the separability 
criterion \cite{Guhne02}. 
In that case, one finds that the optimal witness operator reads:
\begin{equation}
W_{\text{opt}}=\ket{\phi}\bra{\phi}^{\text{T}_{\text{B}}} \;
\end{equation}
where $\ket{\phi}$ is the normalized eigenvector 
corresponding to the negative eigenvalue of the partial 
transpose of the density matrix, i.e. 
$\rho_{\Lambda}^{\text{T}_{\text{B}}}$ ($\text{T}_{\text{B}}$ represents the partial transposition with respect to the subsystem B).

In view of an experimental measurement the mean expectation value of the optimal witness 
$\langle W_{\text{opt}} \rangle$,  it is necessary to decompose
the optimal witness into operators which can 
be measured locally. In this paper, for example, we shall use the 
following decomposition:
\begin{equation}
W_{\text{opt}}=\sum_{n,m=0,x,y,z} \kappa_{nm} \sigma_n \otimes \sigma_m, \label{ciao}
\end{equation}
where $\sigma_0$ is the unity $2\times2$-matrix 
$\openone$ and $\sigma_n$, with ${n=x,y,z}$, are the Pauli matrices.
Notice that, according to this procedure, the average value of the optimal witness operator can be 
calculated {\em only} if the coefficients 
$\kappa_{nm}$ are known. Moreover, it is highly desirable that 
the decomposition given in Eq. (\ref{ciao}) contains a minimal number of terms.
Since we shall only consider optimal witness operators, in the following the word optimal will be understood.

\section{Prototype setup}
\label{ideal}
\begin{figure}
\begin{center}
    \includegraphics[width=0.7 \columnwidth,clip]{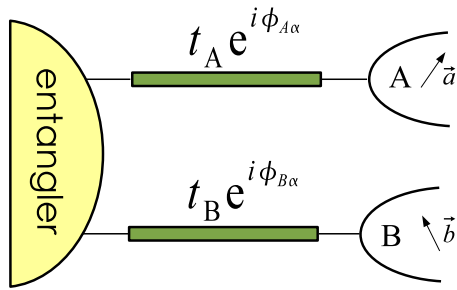}
\end{center}
\caption{Prototype setup consisting of an entangler (yellow region) connected to terminals A and B through two conductors (green regions) characterized by transmission amplitudes $t_{\text{A}}e^{i\phi_{A\alpha}}$ and $t_{\text{B}}e^{i\phi_{B\alpha}}$. The current is measured spin-selectively in A and B along spin quantization axes represented by the vectors $\vec{a}$ and $\vec{b}$.}
\label{Ent}
\end{figure}

For definiteness let us consider first the prototype system presented in Fig.~(\ref{Ent}) composed of an {\em entangler} connected, through two conductors (green wires in the figure), to two outgoing terminals, A and B, where electric current can be measured spin-selectively along specified quantization axis ($\vec{a}$ and $\vec{b}$ for upper and lower electrodes, respectively).
The entangler is a device which produces maximally entangled (Bell) states:
\begin{equation}
\ket{\Psi}=\prod_{0<E<eV} \frac{\left[ a_{\text{A}\uparrow}^{\dagger}(E) a_{\text{B}\downarrow}^{\dagger}(E) +(-1)^s a_{\text{A}\downarrow}^{\dagger}(E) a_{\text{B}\uparrow}^{\dagger}(E) \right]}{\sqrt{2}}\ket{0} ,
\label{ent0}
\end{equation}
where $s=1$ $(0)$ for spin singlet (triplet) and $a_{\text{A}\alpha}^{\dagger}(E)$ ($a_{\text{B}\alpha}^{\dagger}(E)$) is a creation operator for an electron with spin $\alpha$ in the upper (lower) conductor at energy $E$.
The product is taken over energies ranging from the chemical potential of the outgoing terminals (set to zero) and the energy $eV$, while the state $\ket{0}$ represents the filled Fermi sea.
We denote by $t_{\text{A}}$ ($t_{\text{B}}$) the transmission amplitude of the upper (lower) conductor, assuming it as energy-independent in the range $0<E<eV$.
Furthermore, we shall include the presence of ``noise'' in the conductors through two different models.
In the first one (Sec.~\ref{depha}) we only assume dephasing of the spin degree of freedom, while in the second (Sec.~\ref{relax}) we also introduce spin-relaxation.

\subsection{Random phases accumulated in the wires}
\label{depha}

We first consider a rather simplistic model where the 
``noise'' in the transmission wires is simply an acquired random phase.
Specifically, we introduce spin-dependent random phases 
($\phi_{\text{A} \alpha}$ and $\phi_{\text{B} \alpha}$), so that the overall  
effect of the conductors on the electronic operators can be described 
by the transformation:
\begin{equation}
a^{\dagger}_{i \alpha} (E) \to t_i e^{i \phi_{i\alpha}} a^{\dagger}_{i \alpha} (E) ,
\label{Bell2}
\end{equation}
where $i=\text{A},\text{B}$ and $\alpha=\uparrow,\downarrow$.
The density matrix of the outgoing state, i.e. of the electrons arriving at the exit terminals, is given by
\begin{equation}
\rho=\frac{1}{\nu(0)eV} \bigotimes_{0<E<eV} \rho(E)
\label{rho}
\end{equation}
with
\begin{equation}
\rho(E)=\frac{1}{2}\left [ \begin{array}{cccc}
0 & 0 & 0 & 0\\
0 & 1 & (-1)^s  e^{i(x_{\text{A}}-x_{\text{B}})} & 0 \\
0 & (-1)^s e^{-i (x_{\text{A}}-x_{\text{B}})} & 1 & 0 \\
0 & 0 & 0 & 0 
\end{array} \right ]_E ,
\label{ro}
\end{equation}
where ${x_{\text{A}}=\phi_{\text{A} \uparrow}-\phi_{\text{A} \downarrow}}$,  
${x_{\text{B}}=\phi_{\text{B} \uparrow}-\phi_{\text{B} \downarrow}}$ and $\nu(E)$ is the density of states, which we assume to be independent of energy in the range considered.
The subscript $E$ indicates that the matrix refers to the subspace relative to energy $E$.
Note that, as it should be, entanglement is not affected if up and down electrons in the same conductor accumulate an equal random phase, since in that case $x_{\text{A}}=x_{\text{B}}=0$.
We assume a Gaussian distribution for the random phases, i.e.
${P(\phi_{i\alpha})=\frac{1}{\sqrt{2 \pi} \omega} e^{-\frac{\phi_{i\alpha}^2}{2 \omega^2}}}$, with $\omega$ being the width of the distribution, termed {\em dephasing parameter} below.
The phase-averaged density matrix $\langle \rho \rangle$ is given by replacing, in Eq.~(\ref{ro}), $e^{\pm i(x_{\text{A}}-x_{\text{B}})}$ with $e^{-2 \omega^2}$. 

In this setup the two subsystems are identified with conductors A and B, so that the partial transpose with respect to B is given by ${(\rho^{\text{T}_{\text{B}}})_{ij;kl}=(\rho)_{il;kj}}$, where $i$ and $k$ are spin indexes for A, while $l$ and $j$ are spin indexes for B.
In order to construct the witness we perform the partial transpose with respect to the subsystem B on the phase-averaged density matrix and find that the only negative eigenvalue is $\lambda=- \frac{1}{2} e^{-2 \omega^2}$, with eigenvector
\begin{equation}
{\ket{\phi}= \prod_{0<E<eV} \frac{1}{\sqrt{2}} [-a_{\text{A}\uparrow}^{\dagger}(E) a_{\text{B}\uparrow}^{\dagger}(E)+ a_{\text{A}\downarrow}^{\dagger}(E) a_{\text{B}\downarrow}^{\dagger}(E) ]\ket{0}} .
\end{equation}
The fact that $\lambda$ is always negative proves that the outgoing state is entangled for every value of the dephasing parameter $\omega$ smaller than infinity, irrespective of the value of the transmission amplitudes $t_{\text{A}}$ and $t_{\text{B}}$.
The witness is then given by $W= \bigotimes_{0<E<eV} W_E$, where
\begin{eqnarray}
W_E= \frac{1}{2} \left [ \begin{array}{cccc}
1 & 0 & 0 & 0 \\
0 & 0 &(-1)^{s+1} & 0 \\
0 & (-1)^{s+1} & 0 & 0 \\
0 & 0 & 0 & 1 
\end{array} \right ]_E ,
\label{witness}
\end{eqnarray}
so that, by construction, the mean value of the witness operator 
$\langle W \rangle = \text{Tr} (W \rho)$ equals $\lambda$ \cite{nota1}.

It is important to stress that, in order for the witness to be of use, 
$W$ has to be independent of unknown parameters like.
The decomposition, containing the minimal number of terms, for the witness operator in Eq.~(\ref{witness}) is given by \cite{Guhne02}
\begin{equation}
W_E  =\frac{1}{4} [  \openone \otimes \openone  + (-1)^{s+1} \sum_{k=x,y} 
 \sigma_k \otimes \sigma_k + \sigma_z \otimes \sigma_z ] .
\label{dec}
\end{equation}
As it is evident from Eq. (\ref{dec}), operatively the determination of the mean value of $W$ requires the measurement of the
spin correlators ${\langle \vec{a}\cdot\vec{\sigma} \otimes  \vec{b}\cdot\vec{\sigma} \rangle}$ for a few sets of $\vec{a}$ and $\vec{b}$. In the following,
we shall see that under certain conditions the average value of such correlators can be given in terms of current cross-correlators.

To this aim, let us consider the correlator $\langle Q_{A\vec{a}\alpha}(t_{\text{m}}) Q_{B\vec{b}\beta}(t_{\text{m}})\rangle$ \cite{Beenakker05-2,Blanter00}, where ${Q_{A\vec{a}\alpha}(t_{\text{m}})}$ ${\left (Q_{B\vec{b}\beta}(t_{\text{m}}) \right )}$ is the number of particles detected over a measuring time $t_{\text{m}}$ in A (B) with spin $\alpha$ ($\beta$) along the direction $\vec{a}$ ($\vec{b}$).
Intuitively, such a correlator is proportional to a coincidence counting, at least, in the situation where no more than one particle per port arrives at the measuring apparatus within the measuring time $t_{\text{m}}$~\cite{nota3}.
Therefore, one expects the following relation to hold
\cite{Kawabata01,Lesovik01,Chtchelkatchev02,Samuelsson03,Beenakker03}:
\begin{widetext}
\begin{equation}
\langle \vec{a}\cdot\vec{\sigma} \otimes  \vec{b}\cdot\vec{\sigma} \rangle =
\frac{\langle [Q_{A\vec{a}\uparrow}(t_{\text{m}})-Q_{A\vec{a}\downarrow}(t_{\text{m}})]
[Q_{B\vec{b}\uparrow}(t_{\text{m}})-Q_{B\vec{b}\downarrow}(t_{\text{m}})] \rangle}
{\langle [Q_{A\vec{a}\uparrow}(t_{\text{m}})+Q_{A\vec{a}\downarrow}(t_{\text{m}})]
[Q_{B\vec{b}\uparrow}(t_{\text{m}})+Q_{B\vec{b}\downarrow}(t_{\text{m}})] \rangle} ,
\label{spin1}
\end{equation}
\end{widetext}
with correlators defined as \cite{nota2}
\begin{equation}
\langle Q_{A\vec{a}\alpha}(t_{\text{m}}) Q_{B\vec{b}\beta}(t_{\text{m}})\rangle=
\int_0^{t_{\text{m}}} dt \int_0^{t_{\text{m}}} dt' \langle I^{\vec{a}}_{A\alpha}(t) I^{\vec{b}}_{B\beta}(t') \rangle .
\label{qq}
\end{equation}
In Eq.~(\ref{qq}) $I^{\vec{a}}_{A\alpha}(t)$ is the current operator, defined in Appendix \ref{app:pro}, for electrons arriving at terminal A with 
spin $\alpha$ along the direction $\vec{a}$, 
and analogously for $I^{\vec{b}}_{B\beta}(t')$.
We now make use of the fact that the correlator in Eq.~(\ref{qq}) can be expressed in terms of the {\em zero-frequency current-correlator} defined as
\begin{equation}
s_{\alpha \beta}^{\vec{a}\vec{b}}(0) \equiv  \int_{-\infty}^{\infty} dt \langle
\delta I^{\vec{a}}_{A\alpha}(t) 
\delta I^{\vec{b}}_{B\beta} (0) \rangle ,
\label{cross}
\end{equation} 
with ${\delta I^{\vec{a}}_{j \alpha }(t)= I^{\vec{a}}_{j \alpha}(t)- \langle I^{\vec{a}}_{j \alpha } \rangle}$.
The actual expression depends on whether the measuring time $t_{\text{m}}$ is larger or smaller than the correlation time $\tau_{\text{c}}=h/eV$, which represents the time spread of the electron wave packet.
Namely, for $t_{\text{m}}\gg \tau_{\text{c}}$ one gets \cite{Beenakker05-2}
\begin{equation}
\langle Q_{A\vec{a}\alpha}(t_{\text{m}}) Q_{B\vec{b}\beta}(t_{\text{m}})\rangle \simeq
t_{\text{m}}^2 \langle I^{\vec{a}}_{A\alpha} \rangle \langle I^{\vec{b}}_{B\beta} \rangle +
t_{\text{m}} s_{\alpha \beta}^{\vec{a}\vec{b}}(0)
\end{equation} 
and, for $t_{\text{m}}\ll \tau_{\text{c}}$, one gets
\begin{equation}
\langle Q_{A\vec{a}\alpha}(t_{\text{m}}) Q_{B\vec{b}\beta}(t_{\text{m}})\rangle \simeq
t_{\text{m}}^2 [\langle I^{\vec{a}}_{A\alpha} \rangle \langle I^{\vec{b}}_{B\beta} \rangle +
\frac{eV}{h} s_{\alpha \beta}^{\vec{a}\vec{b}}(0)] .
\end{equation}
At zero temperature and for short measuring times ($t_{\text{m}}\ll \tau_{\text{c}}$), the spin correlator for our setup turns out to be
\cite{footnoteNo2}
\begin{equation}
\langle \vec{a}\cdot\vec{\sigma} \otimes  \vec{b}\cdot\vec{\sigma} \rangle =
\frac{eV}{2h}\frac{s_{\uparrow \uparrow}^{\vec{a}\vec{b}}(0)+s_{\downarrow \downarrow}^{\vec{a}\vec{b}}(0)-
s_{\uparrow \downarrow}^{\vec{a}\vec{b}}(0)-s_{\downarrow \uparrow}^{\vec{a}\vec{b}}(0)}
{\langle I^{\vec{a}}_{A\uparrow}+ I^{\vec{a}}_{A\downarrow}  \rangle 
\langle I^{\vec{b}}_{B\uparrow}+ I^{\vec{b}}_{B\downarrow}  \rangle} .
\label{spin}
\end{equation}
By using Eq.~(\ref{dec}), we find that the mean value of the witness is
\begin{equation}
\langle W \rangle =  -\frac{1}{2}  e^{-2 \omega^2} ,
\end{equation}
which equals the values found for $\lambda$.
On the contrary, in the limit of large measuring times ($t_{\text{m}}\gg \tau_{\text{c}}$) we obtain:
\begin{equation}
\langle \vec{a}\cdot\vec{\sigma} \otimes  \vec{b}\cdot\vec{\sigma} \rangle =
\frac{1}{2t_{\text{m}}}\frac{s_{\uparrow \uparrow}^{\vec{a}\vec{b}}(0)+s_{\downarrow \downarrow}^{\vec{a}\vec{b}}(0)-
s_{\uparrow \downarrow}^{\vec{a}\vec{b}}(0)-s_{\downarrow \uparrow}^{\vec{a}\vec{b}}(0)}
{\langle I^{\vec{a}}_{A\uparrow}+ I^{\vec{a}}_{A\downarrow}\rangle 
\langle I^{\vec{b}}_{B\uparrow}+ I^{\vec{b}}_{B\downarrow}\rangle} ,
\label{spinno}
\end{equation}
which is always vanishing, even in the tunneling regime.

This is due to the fact that, in our setup, current cross-correlations of Eq.~(\ref{cross}) are inherently a result of two particle scattering processes.
More precisely, current correlations are proportional to $|t_{\text{A}}|^2 |t_{\text{B}}|^2$ likewise the product $\langle I^{\vec{a}}_{A\alpha}\rangle \langle I^{\vec{b}}_{B\beta}\rangle$.
In contrast, in Refs.~\onlinecite{Samuelsson03,Beenakker03}, for example, the cross-correlations result from partition noise and, in the tunneling regime, are much larger than the 
product of the currents (times $t_{\text{m}}$) 
provided that the measuring time is smaller than the average time 
spacing between tunneling events.
It results that for large measuring times one finds 
$\langle W \rangle \simeq \frac{1}{4}$, which would be the result obtained
averaging over a fully mixed state.
We conclude that the witness operator can be measured only for measuring 
times $t_{\text{m}}$ much smaller than the correlation time $\tau_{\text{c}}$.
Nevertheless, entanglement can be detected even beyond the latter condition.
Indeed, for arbitrary measuring times $t_{\text{m}}$ the spin correlator can be written as:
\begin{equation}
\langle \vec{a}\cdot\vec{\sigma} \otimes  \vec{b}\cdot\vec{\sigma} \rangle =
\alpha (t_{\text{m}},\tau_{\text{c}})
\frac{s_{\uparrow \uparrow}^{\vec{a}\vec{b}}(0)+s_{\downarrow \downarrow}^{\vec{a}\vec{b}}(0)-
s_{\uparrow \downarrow}^{\vec{a}\vec{b}}(0)-s_{\downarrow \uparrow}^{\vec{a}\vec{b}}(0)}
{\langle I^{\vec{a}}_{A\uparrow}+ I^{\vec{a}}_{A\downarrow}  \rangle 
\langle I^{\vec{b}}_{B\uparrow}+ I^{\vec{b}}_{B\downarrow}  \rangle} ,
\label{qua}
\end{equation}
where $\alpha$ is, in general, a function of both $\tau_{\text{c}}$ and $t_{\text{m}}$.
The following simple relation can then be derived:
\begin{equation}
\langle W\rangle=2\alpha(t_{\text{m}},\tau_{\text{c}}) \tau_{\text{c}}\lambda+\frac{1}{4}\left[ 1- 2\alpha(t_{\text{m}},\tau_{\text{c}}) \tau_{\text{c}} \right] .
\end{equation}
It turns out that, for $2\alpha\tau_{\text{c}}\geq 1$, $\langle W\rangle <0$ implies $\lambda<0$ so that  entanglement can be witnessed as long as $\langle W\rangle$ is negative.
Furthermore, since the spin correlator in Eq. (\ref{qua}) depends explicitly on $t_{\text{m}}$, its measurement requires time-resolved detection, i. e. energy filtering. This is not the case in the limit $t_{\text{m}} \ll \tau_{\text{c}}$.

It is interesting to compare the capability of detection of entanglement provided by the witness with the one relative to a Bell inequality test.
In the following we choose the well known Clauser-Horne-Shimony-Holt (CHSH) inequality
\cite{Kawabata01,Lesovik01,Chtchelkatchev02,Samuelsson03,Beenakker03}:
\begin{equation}
{\cal E}=\left| E(\vec{a},\vec{b})+ E(\vec{a}',\vec{b})+E(\vec{a}',\vec{b})-E(\vec{a}',\vec{b}') \right | \le 2,
\label{bellpar}
\end{equation}
where $E(\vec{a},\vec{b}) = \langle \vec{a}\cdot\vec{\sigma} \otimes  
\vec{b}\cdot\vec{\sigma} \rangle$
is the spin correlator given in Eq.~(\ref{spin}) and 
${\vec{a},\vec{a}',\vec{b},\vec{b}'}$ denote 
arbitrary directions along which the spin is measured.
In the absence of dephasing, for the case of the 
Bell state given in Eq.(\ref{ent0}), the CHSH inequality is maximally violated by choosing the following set of directions:
\begin{eqnarray}
\vec{a}&=&(1,0,0) ~~~~~~ \vec{b}=\frac{1}{\sqrt{2}}(1,0,1) \notag \\
\vec{a}'&=&(0,0,1) ~~~~~~ \vec{b}'=\frac{1}{\sqrt{2}}(1,0,-1).
\label{set}
\end{eqnarray}
Indeed one finds that 
${\cal E}={\cal E}_{\text{max}}\equiv 2 \sqrt{2}$ and, as a result, the entanglement of the Bell state is detected.
Notice that, in order to experimentally test the inequality, for every set of directions one needs to measure $4$ spin correlators and, for each of them, $4$ current cross-correlators.

Let us now assume the presence of dephasing in the wires. It is easy to verify that, for ${t_{\text{m}}\ll \tau_{\text{c}}}$, 
${\cal E}=\sqrt{2}(1+e^{-2 \omega^2})$. As a result, for value of the 
dephasing 
parameter ${\omega>\sqrt{\frac{1}{2} |\ln (\sqrt{2}-1)|}}$ the Bell inequality is not violated even thought, according to the previous witness 
detection analysis, the state exiting the leads is still entangled.

It has to be noticed, however, that if the dephasing parameter $\omega$ is known, it is possible to find the proper set of directions leading 
to violation of the CHSH inequality.
In practice, since the value of $\omega$ is unknown, in order to find a 
violation of the inequality, one needs to try several measurements of 
current cross-correlators over different sets of directions.  
On the contrary, by measuring the witness given in Eq. (\ref{dec}) 
one needs to perform only $12$ measurements of current cross-correlators, 
$4$ for each spin correlator, with fixed configurations irrespective of the 
degree of dephasing in the conductors.

As a final remark, we notice that the measurement of the mean value of the 
witness gives information on the degree of the dephasing affecting the wires.

\subsection{Spin relaxation and dephasing}
\label{relax}

We shall here analyze the same system considered in the previous section, now
assuming that during the propagation in the wires the electron spins are 
subject to relaxation in addition to dephasing.
The former might be caused by the presence of magnetic impurities, 
nuclear spins, or the spin-orbit coupling in the wires \cite{Awschalom}.
We shall adopt a phenomenological model of decoherence 
along the lines of Ref.~\onlinecite{Burkard00}, by resorting to the Bloch theory \cite{Bloch} and by expressing the density matrix at 
each exit leads as: 
\begin{equation}
\rho_i(t_L)\equiv \Lambda_i(t_L)[\rho_i(0)]
\end{equation}
with $i=\text{A},\text{B}$.
Here ${t_L=L/v_{\text{F}}}$ is the ballistic transmission time 
($L$ is the length of the wires and $v_{\text{F}}$ is the Fermi velocity) and we 
have introduced the superoperator:
\begin{widetext}
\begin{eqnarray}
\Lambda_i(t_L)[\rho] =\frac{1}{2}  \left [ \begin{array}{cccc}
\rho_{\uparrow \uparrow} \left (1+  e^{-t_L/ T_1} \right ) + 
\rho_{\downarrow \downarrow} \left (1- e^{-t_L/T_1} \right ) 
& 2 \, e^{-t_L/T_2} \rho_{\uparrow \downarrow} \\
2 \, e^{-t_L/T_2} \rho_{\downarrow \uparrow} &  \rho_{\uparrow \uparrow} \left (1-  e^{-t_L/T_1} \right ) + 
\rho_{\downarrow \downarrow} \left (1 + e^{-t_L/T_1} \right ) 
\end{array} \right ] ,
\end{eqnarray}
\end{widetext}
where $T_1$ is the relaxation time and $T_2$ is the dephasing time.
By assuming that the decoherence and relaxation processes act on each wire 
independently, we can readily write the density matrix of the outgoing 
state as follows:
\begin{eqnarray}
\rho(t_L) = \left (\Lambda_\text{A}(t_L) \otimes \Lambda_\text{B} (t_L) 
\right ) [\rho(0)] \;
\label{dens}
\end{eqnarray}
where $\rho(0)$ is the density matrix of the initial entangled state given in Eq. (\ref{ent0}).
It is a matter of simple calculations to show that
\begin{eqnarray}
\rho(t_L) =\frac{1}{4}  \left [ \begin{array}{cccc}
d_- & 0 & 0 & 0 \\
0 &  d_+ &  (-1)^s 2\;e^{-2 t_L/T_2} \\
0 &  (-1)^s 2\; e^{-2 t_L/T_2} &  d_+   & 0 \\
0 & 0 & 0 &  d_-
\end{array} \right ], \notag
\end{eqnarray}
where ${d_\pm = \left (1 \pm e^{-2 t_L/ T_1} \right )}$.
\begin{figure}[t!]
    \begin{center}
    \includegraphics[width=0.9 \columnwidth,clip]{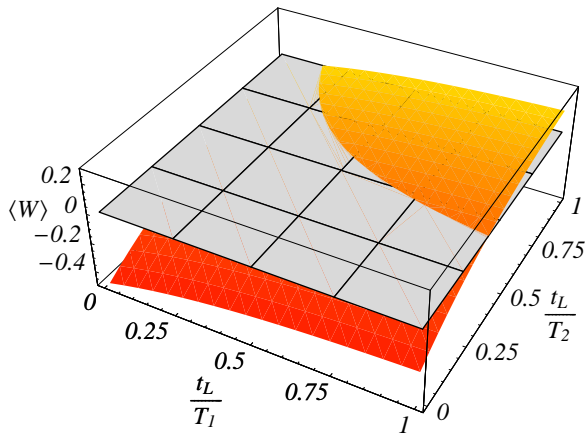}
    \end{center}
    \caption{(Color online) Mean value of the witness as a function of inverse relaxation ($t_L/T_1$) and dephasing ($t_L/T_2$) times. The gray plane corresponds to $\langle W\rangle=0$. A typical GaAs structure has 
${L\approx 1 \mu m}$ and ${v_F \approx 10^4 - 10^5 m/s}$, so that
we obtain ${t_L \approx 10-100 ps}$. Typical decoherence time found in 
GaAs are of the order ${T_2 \approx 100 ns-1 \mu s}$. It follows that 
${t_L/T_2 \approx 10^{-5}-10^{-3}}$. Typically $T_1 \gg T_2$.}
    \label{witT1T2}
\end{figure}
The witness operator can now be determined by applying the method described in Sec.~\ref{depha}.
We find that it is equal to the one reported in Eq.~(\ref{witness}) and its mean value reads:
\begin{equation}
\langle W \rangle \equiv \text{Tr} \bigl (W\rho(t_L) \bigr ) = \frac{1}{4} \left ( 1-e^{-2 \frac{t_L}{T_1}} - 2 e^{-2 \frac{t_L}{T_2}} \right ).
\label{lamT1T2}
\end{equation}
In Fig.~\ref{witT1T2} we report a three-dimensional plot of 
${\langle W \rangle}$ as a function of $t_L/T_1$ and $t_L/T_2$.
To highlight the region of the surface which corresponds to entanglement, we have plotted a gray plane in correspondence to $\langle W \rangle=0$.
Entanglement is present when the orange surface is below the gray plane.
For large relaxation times ($t_L/T_1 \to 0$) we find that 
${\langle W \rangle}$ is always negative: in agreement with 
the results of Sec.~\ref{depha}, we find that for arbitrary degrees of 
dephasing there is always entanglement in the outgoing states. 
However,  for finite relaxation time, there always exists a threshold  
for the dephasing above which entanglement is lost during the 
transmission in the wires.  Indeed 
it is easy to verify that for ${t_L/T_2 > \frac{1}{2} |\ln \frac{1}{2} \left (1- e^{-2 t_L/T_1} \right )|}$ the mean value of the witness becomes 
positive. Note furthermore that relaxation processes have a much less disrupting effect with respect to dephasing processes, due to the prefactor 2 in Eq.~(\ref{lamT1T2}).

\begin{figure}[t!]
    \begin{center}
    \includegraphics[width=0.9 \columnwidth,clip]{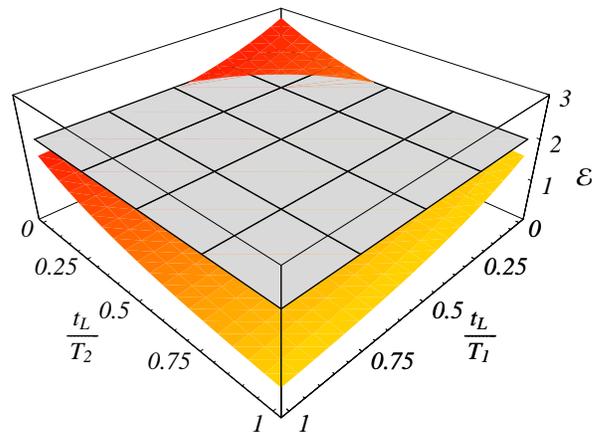}
    \end{center}
    \caption{(Color online) CHSH parameter ${\cal E}$ as a function of inverse relaxation ($t_L/T_1$) and dephasing ($t_L/T_2$) times. The gray plane corresponds to ${\cal E}=2$.}
    \label{emax}
\end{figure}

Let us again compare the performance of the witness to the Bell inequality test. 
We find that, choosing the set of directions given in Eq.~(\ref{set}), the CHSH parameter reads
\begin{equation}
{\cal E}=|\sqrt{2}\left( e^{-2 \frac{t_L}{T_1}}+ e^{-2 \frac{t_L}{T_2}} \right)| .
\end{equation}
In Fig.~\ref{emax} we plot the CHSH parameter as a function of ${t_L/T_1}$ and 
${t_L/T_2}$.
We draw a gray plane in correspondence to ${\cal E}=2$ in order to highlight the region in the parameter space (above the plane) for which entanglement is detected. 
By comparison with Fig.~\ref{witT1T2}, it is evident that there is a large region in the parameter space for which Bell inequality fails to detect entanglement.
Notice that also in this case ${\cal E}$ can be maximized by exploring different sets of angles.

\section{Particle-hole entanglement}
\label{ph}

\subsection{Edge states in a quantum Hall bar with a quantum point contact}
\label{Hall}

\begin{figure}[h]
    \begin{center}
    \includegraphics[width=0.9 \columnwidth,clip]{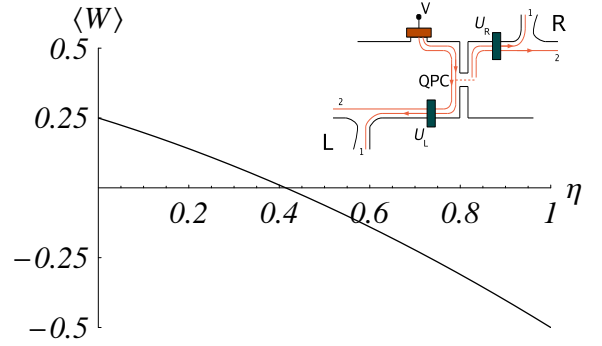}
    \end{center}
    \caption{Witness mean value as a function of the dephasing parameter $\eta$ for particle-hole entanglement. Inset: the system consists of a Hall bar divided in two halves by a QPC. A large magnetic filed is applied perpendicular to the two dimensional electron gas, so that it allows two edge states labeled by 1 and 2.
Electrons are injected from the contact biased at a voltage $V$, red rectangle, and are transmitted to the right-hand-side or reflected to the left-hand-side. Two additional gate electrodes, represented by blocks $U_{\text{L}}$ and $U_{\text{R}}$, are placed on both sides of the bar and provide a controlled mixing of the two channels. Electric current is then measures at the exit ports L and R.}
    \label{Hallfig}
\end{figure}

It is interesting to apply the method of the witness to some realistic systems.
Let us now consider the particle-hole entanglement of edge channels in the quantum Hall bar system proposed in Ref.~\onlinecite{Beenakker03}.
The structure, depicted in the inset of Fig.~\ref{Hallfig}, consists of a quantum Hall bar divided in two halves by a quantum point contact (QPC).
A large magnetic field is applied perpendicular to the plane, so that two edge states are allowed.
Electrons are injected from a contact, represented by a red rectangle, biased at a voltage $V$.
They propagate to the right through the two edge channels (labeled by 1 and 2), and are transmitted to the right-hand-side with amplitudes matrix $t$ and reflected in the left-hand-side with amplitude matrix $r$.
In the tunneling limit, the outgoing state can be written as \cite{Beenakker03}
\begin{equation}
\ket{\Psi}=\prod_E\left( \sqrt{\Omega}\ket{\bar{\Psi}}+\sqrt{1-\Omega} \ket{\bar{0}}\right) ,
\label{beeen}
\end{equation}
where
\begin{equation}
\ket{\bar{\Psi}}=\Omega^{-1/2}\underline{b}_{\text{h}}^{\dagger}\gamma \underline{b}_{\text{p}}^{\dagger} \ket{\bar{0}}
\label{been}
\end{equation}
is an entangled state expressed in terms of particle operators $\underline{b}_{\text{p}}^{\dagger}=(b_{\text{p}1}^{\dagger},b_{\text{p}2}^{\dagger})$ for the right-hand-side and hole operators $\underline{b}_{\text{h}}^{\dagger}=(b_{\text{h}1}^{\dagger},b_{\text{h}2}^{\dagger})$ for the left-hand-side.
In Eqs.~(\ref{beeen}) and (\ref{been}), $\Omega=\text{Tr}\gamma \gamma^{\dagger}$, $\gamma=\sigma_y r \sigma_y t^{\text{T}}$ and $\ket{\bar{0}}$ represents a new vacuum state consisting of particle-filled edge states in the left-hand-side.
Two additional gate electrodes, which provide a local mixing of the channels, are placed on the left and on the right with scattering matrix $U_{\text{L}}$ and $U_{\text{R}}$, respectively.
Current cross-correlators, which determine the spin correlator through Eq.~(\ref{spin1}), are measured between the exit ports L and R.

Let us consider the following parametrization of $\gamma$ \cite{beenakker03-3}:
\begin{equation}
\gamma=e^{i\theta}\sqrt{{\cal T}(1-{\cal T})}
\left( \begin{array}{cc} e^{i(\phi_{\text{L}2}+\phi_{\text{R}1})}\cos{\xi} & e^{i(\phi_{\text{L}2}+\phi_{\text{R}2})}\sin{\xi} \\ -e^{i(\phi_{\text{L}1}+\phi_{\text{R}1})}\sin{\xi} & e^{i(\phi_{\text{L}1}+\phi_{\text{R}2})}\cos{\xi} \end{array}\right)
\end{equation}
which implies identical transmission eigenvalues ${\cal T}$ for the two channels.
The possible presence of dephasing processes is accounted for by introducing Gaussian-distributed random phases ${\phi_{\text{L(R)}i}}$ with equal width $\omega$.
$\xi$ is the channel-mixing parameter of the QPC.
We find one negative eigenvalue 
\begin{equation}
\lambda=\frac{\sqrt{2}(1-\eta^2)-\sqrt{1+6\eta^2+\eta^4+(\eta^2-1)^2 \cos{4\xi}}}{4\sqrt{2}}
\label{lambee}
\end{equation}
for the partially transposed phase-averaged density matrix, where ${\eta=e^{-\omega^2}}$, showing that entanglement exists only within some ranges of the parameters $\xi$ and $\eta$.
It is interesting to note that $\lambda$ is related to the concurrence ${\cal C}$ calculated in Ref.~\onlinecite{beenakker03-3}, namely $\lambda=-{\cal C}/2$ for positive values of ${\cal C}$.

The witness operator, constructed with the separability criterion as in Sec.~\ref{ideal}, in this case depends on such parameters and cannot be used to detect entanglement in the general case.
If, however, the mixing parameter $\xi$ is known or controllable, then a witness independent of parameters can be constructed for maximum ($\xi=\pi/4$) or no mixing ($\xi=0,\pi/2$).
In this latter case $W_E$ takes the form of Eq.~(\ref{dec}), while in the former we find
\begin{equation}
W_E =\frac{1}{4} \left \{  \openone \otimes \openone  + 
 \sigma_y \otimes \sigma_y  -  \sigma_z \otimes \sigma_x  +   \sigma_x \otimes \sigma_z  \right \} .
\label{decbeen}
\end{equation}

Even for this system the possibility of measuring the witness depends on the relative position of the time scales involved: measuring time $t_{\text{m}}$, correlation time $\tau_{\text{c}}=h/eV$ and tunneling time $\tau_{\text{tun}}=h/(2eV{\cal T})$. Namely, both in the maximum and minimum mixing conditions, for $t_{\text{m}}\gg \tau_{\text{c}}$ we find that the mean value $\langle W\rangle$ is measurable and equals the value of $\lambda$ given in Eq.~(\ref{lambee}) only in the tunneling regime ($t_{\text{m}}\ll \tau_{\text{tun}}$).
Nonetheless, for $t_{\text{m}}\ll \tau_{\text{c}}$ the witness can be measured irrespective of the value of ${\cal T}$.
Specifically, in the case of no mixing, we find that 
$\langle W\rangle=-\eta^2/2$, showing that, 
irrespective of the degrees of the dephasing, 
the state is always entangled. 
Interestingly, this result equals the 
witness mean value obtained in Sec.~\ref{depha}.
On the contrary, in the case of maximum 
mixing we find that $\langle W\rangle=1/4(1-2\eta-\eta^2)$. We plot this 
result in Fig.~\ref{Hallfig} as a function of $\eta$. Notice that
the state is entangled only for values of 
${\eta>\sqrt{3-2\sqrt{2}}}\simeq 0.4$. 
This result can be compared with the prediction given by a 
Bell inequality test that assures that entanglement is present only for 
${\eta>1/\sqrt{2}\simeq 0.7}$ (see the discussion in Ref.~\onlinecite{beenakker03-3}).
Unfortunately, no parameter-free witness could be found for arbitrary mixing $\xi$.

\subsection{Hanbury-Brown and Twiss interferometer}
\label{HBT}
We shall now demonstrate that it is possible to 
use the witness method to detect the orbital entanglement generated in the electronic equivalent of the optical Hanbury-Brown and Twiss (HBT) experiment, proposed in Ref.~\onlinecite{Samuelsson04}.
\begin{figure}[h]
    \begin{center}
    \includegraphics[width=0.8 \columnwidth,clip]{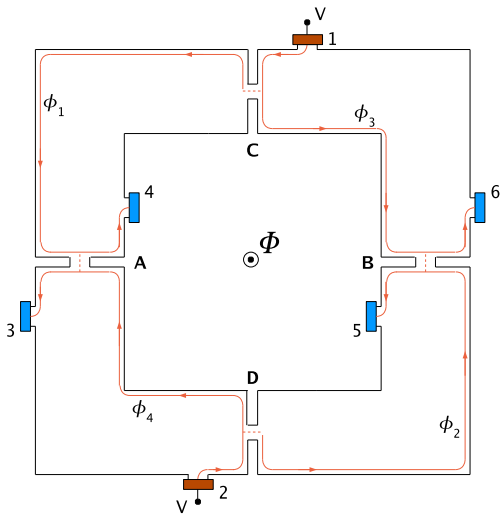}
    \end{center}
    \caption{Sketch of the electric Hanbury Brown-Twiss interferometer. The edges of the Hall bar are drawn with black lines, edges states are represented by red lines and a magnetic flux $\Phi$ threads through the center of the structure. The four QPCs are labeled with the letters A, B, C, and D, while the six contacts are depicted by colored rectangles. Contacts 1 and 2 are biased at a voltage $V$, while all the other are grounded. Electrons, propagating along the edge states, accumulate phases $\phi_1$, $\phi_2$, $\phi_3$ and $\phi_4$.}
    \label{HBTfig}
\end{figure}
The electronic HBT interferometer, sketched in Fig.~\ref{HBTfig}, consists of a rectangular Hall bar, whose inner and outer edges are indicated with black lines, and four QPCs, labeled with A, B, C and D.
Six contacts are present, indexed with a number from 1 to 6.
Electrons are injected from sources $1$ and $2$, where 
a voltage $V$ is applied against all other contacts which are grounded. 
Each QPC acts as a beam splitter and can be parametrized by the following scattering matrix:
\begin{eqnarray}
G_i =\left ( \begin{array}{cc}
\cos \theta_i e^{i \varphi^i_{11}}& \sin \theta_i e^{i \varphi^i_{12}} \\
\sin \theta_i e^{i \varphi^i_{21}} & \cos \theta_i e^{i \varphi^i_{22}}
\end{array} \right ) .
\label{bs}
\end{eqnarray}
At each QPC electrons are transmitted with probability ${\cal T}_i\equiv \cos^2 \theta_i$ (with $i=$A,B,C,D) and reflected with probability ${\cal R}_i$, with ${{\cal R}_i=1-{\cal T}_i}$.
The phases $\varphi^i_{jk}$ are real and satisfy the condition 
${\varphi^i_{11}+ \varphi^i_{22}=\pi + \varphi^i_{12}+ \varphi^i_{21}}$ 
in order to guarantee the unitarity of the matrix $G_i$, i.e. ${G_i^\dagger G_i = \openone}$.
Along the bar, the electrons follow the edge states (red lines) in 
the directions indicated by the arrows and accumulate phases 
$\phi_1$ to $\phi_4$ (see Fig.~\ref{HBTfig}).
They can be modulated by varying the length of the paths through additional gates.
Since there are no interfering orbits, no single-particle interference will take place and conductance measured at contacts 3 to 6 will be phase-insensitive.
Because of the indistinguishability of the paths taken by the particles injected from sources $1$ and $2$, two-particle interference effects arise influencing the current cross-correlations measured at the grounded contacts.

Let us briefly recall here the entangled state generated in this system by closely following Ref.~\onlinecite{Samuelsson04}.
To this aim, it is useful to denote the transmission and reflection probabilities at the QPC C with ${{\cal T}_{\text{C}}=1-{\cal R}_{\text{C}}={\cal T}}$ and at the QPC D with ${{\cal T}_{\text{D}}=1-{\cal R}_{\text{D}}={\cal R}}=1-{\cal T}$. 
The state generated by the two independent sources $1$ and $2$ can be written as: 
${|\Psi \rangle = \prod_{0<E<eV} a^{\dagger}_1(E)a^\dagger_2(E)|0\rangle}$ 
where $|0\rangle$ is the ground state, describing a filled Fermi sea in all reservoirs 
at energies $E<0$ and we have omitted the spin index. The operator $a^\dagger_j(E)$ creates an injected electron from reservoir $j$ at energy $E$. 
In the case of strong asymmetry, i.e. ${{\cal R} \ll 1}$, it is possible to 
express the state ${|\Psi \rangle}$ to leading order in ${\sqrt{{\cal R}}}$ as
\begin{equation}
|\Psi\rangle = |\bar{0}\rangle + \sqrt{{\cal R}} |\tilde{\Psi}\rangle  ,
\label{initial}
\end{equation}
where
\begin{equation}
|\tilde{\Psi}\rangle=\int_0^{eV} dE \left [b^\dagger_{2\text{B}} (E) b_{2\text{A}}(E)-b^\dagger_{1\text{B}} (E) b_{1\text{A}}(E) \right ] |\bar{0}\rangle
\label{minchia}
\end{equation}
and $b^{\dagger}_{1i}$ ($b^{\dagger}_{2i}$) is the creation operator for electrons coming from contact 1 (2) and exiting from the QPCs C and D. 
The second index in the operator indicates towards which QPC the electrons are propagating.
The state
\begin{equation}
|\bar{0}\rangle = \prod_{0<E<eV} b^\dagger_{1\text{A}} (E) b^\dagger_{2\text{A}}(E) |0 \rangle
\end{equation}
can be thought as a new vacuum state representing a completely filled stream of electrons.
The state ${|\tilde{\Psi}\rangle}$ describes an orbitally entangled pair of electron-hole excitations \cite{Beenakker03}. 

In order to simplify the witness construction it is useful to re-write 
the electron-hole entangled state of Eq.~(\ref{minchia}) in terms of the vacuum state ${|0\rangle}$:
\begin{equation}
|\tilde{\Psi}\rangle=\prod_{0<E<eV} \left [b^{\dagger}_{1\text{A}}(E) b^{\dagger}_{2\text{B}}(E)+ b^{\dagger}_{2\text{A}}(E) b^{\dagger}_{1\text{B}}(E)\right ]|0\rangle
\label{entbut}
\end{equation}

Let us now assume that the HBT is subject to dephasing. Specifically, we
consider that the electrons propagating along the wires besides the phases 
$\phi_k$ accumulate a random phase $x_k$, with ${k=1, ..., 4}$.  
Moreover, this can include a randomly fluctuating external magnetic flux $\Phi$.
Formally, we introduce the dephasing as follows:
\begin{eqnarray}
b^{\dagger}_{1 \text{A}} (E) &\to& \sqrt{{\cal T}} e^{i (\phi_1 + x_1)} b^{\dagger}_{1 \text{A}} (E)\;\notag \\
b^{\dagger}_{2 \text{A}} (E) &\to& \sqrt{{\cal T}} e^{i (\phi_4 + x_4)} b^{\dagger}_{2 \text{A}} (E)\notag \\
b^{\dagger}_{1 \text{B}} (E)&\to& \sqrt{{\cal R}} e^{i (\phi_3 + x_3)} b^{\dagger}_{1 \text{B}} (E) \; \notag \\
b^{\dagger}_{2 \text{B}} (E) &\to& \sqrt{{\cal R}} e^{i (\phi_2 + x_2)} ~b^{\dagger}_{2 \text{B}} (E) \; .
\notag
\end{eqnarray}
It is quite straightforward then to calculate the density matrix $\rho$ of the 
outgoing entangled state $\ket{\tilde{\Psi}}$:
\begin{equation}
\rho=\frac{1}{\nu(0)eV} \bigotimes_{0<E<eV} \rho(E)
\label{rhoHBT}
\end{equation}
with
\begin{equation}
\rho(E)=\left [ \begin{array}{cccc}
0 & 0 & 0 & 0\\
0 & 1 & e^{-i \phi_0} e^{-i \delta} & 0 \\
0 & e^{i \delta} e^{i \phi_0} & 1 & 0 \\
0 & 0 & 0 & 0 
\end{array} \right ]_E ,
\label{robut}
\end{equation}
where we have introduced the total phase ${\phi_0=\phi_1+\phi_2-\phi_3-\phi_4 + 2 \pi \Phi/\Phi_0}$ and the dephasing parameters ${\delta=x_1+x_2-x_3-x_4}$.
By assuming Gaussian distributions for the random phases, we take into account the presence of dephasing by replacing $e^{i\delta}$ with its average value $\langle e^{-i \delta} \rangle = e^{- \omega^2}$ in the density matrix $\rho$.

We can then proceed to calculate the witness operator $W=\bigotimes_{0<E<eV} W_E$ finding that
\begin{eqnarray}
W_E= \frac{1}{2} \left [ \begin{array}{cccc}
1 & 0 & 0 & 0 \\
0 & 0 &- e^{i \phi_0} & 0 \\
0 & - e^{-i \phi_0} & 0 & 0 \\
0 & 0 & 0 & 1 
\end{array} \right ]_E ,
\label{witnessbut}
\end{eqnarray}
whose mean value correctly gives $\langle W \rangle = Tr (\rho W)=-e^{- \omega^2}$.
We conclude, on the one hand, that the entangled electron-hole pair is robust against arbitrary degrees of 
dephasing in the HBT and, on the other, that the measurement of the witness gives an estimate of the degree of dephasing afflicting the HBT.

The witness can be decomposed as follows:
\begin{eqnarray}\notag
W_E=\frac{1}{4} \left [ \langle \openone \otimes \openone \rangle + \langle \sigma_z \otimes \sigma_z \rangle -
\right. \\ \left. -
\cos \phi_0 \left ( \langle \sigma_x \otimes \sigma_x \rangle+
\langle \sigma_y \otimes \sigma_y \rangle \right ) -
\right. \\ \left.
- \sin \phi_0 \left ( \langle \sigma_x \otimes \sigma_y \rangle + \langle \sigma_y \otimes \sigma_x \rangle \right ) \right ] . \notag
\label{odeccbut}
\end{eqnarray}
The witness can  be evaluated by measuring the zero 
frequency current cross-correlations $s^{\vec{a}\vec{b}}_{\alpha\beta}(0)$ 
of the current fluctuations $\delta I_\alpha$ and $\delta I_\beta$ 
at contacts  ${\alpha \in 3,4}$ and ${\beta \in 5,6}$ (see App.~\ref{app:hbt}).
In fact, although the state exiting the QPCs C and D is the one given in Eq.~(\ref{initial}), it is easy to realize that 
the only part that contributes to cross-correlation measurements at 
QPCs A and B is indeed the entangled electron-hole pair state 
${|\tilde{\Psi}\rangle}$. The vacuum state $|\bar{0}\rangle$, indeed,
carries a current from sources to the detectors but does not 
contribute to the cross-correlators.  

At zero temperature and in the limit of long measuring times  ${\tau_c\ll t_m \ll \tau_{\text{tun}}}$, the spin correlators appearing in Eq.~(\ref{odeccbut}) can be expressed as
\begin{equation}
\langle \vec{a}\cdot\vec{\sigma} \otimes \vec{b}\cdot\vec{\sigma} \rangle = \frac{h}{2 e^3 V {\cal RT}} \left (  s_{36}^{\vec{a}\vec{b}}(0)+s_{45}^{\vec{a}\vec{b}}(0) - s_{35}^{\vec{a}\vec{b}}(0)- s_{46}^{\vec{a}\vec{b}}(0) \right ) .
\end{equation}
The expressions for the different values of $s_{\alpha\beta}^{\vec{a}\vec{b}}(0)$ are reported in Appendix~\ref{app:hbt}.
Here we have used the fact that in the HBT interferometer the average total currents at the QPCs A and B are 
${\langle I_3 \rangle=\langle I_4 \rangle = \frac{e^2}{h}{\cal T} V}$ and
${\langle I_5 \rangle=\langle I_6 \rangle = \frac{e^2}{h}{\cal R} V}$ \cite{Samuelsson04}.  
Finally we find
\begin{equation}
\langle W \rangle = -\cos \phi_0 \langle \cos (\phi_0+\delta) \rangle - \sin \phi_0 \langle \sin (\phi_0+\delta) \rangle = -e^{-\omega^2} 
\end{equation}
where the last equality is obtained by averaging the variable $\delta$ over the Gaussian distribution.

\section{Conclusions}
\label{conc}
In this paper we discussed how the concept of entanglement witness
can be used in order to detect entanglement for electrons in multiterminal 
mesoscopic conductors afflicted by ``noise''. 
By resorting to the separability criterion, we constructed an
optimal witness operator for a prototype setup where maximally entangled spin 
singlet and triplet states are generated.
It has been shown that such witness operator can be decomposed into a sum of local spin correlators and, in the limit of measuring time much smaller than the electron 
correlation time, its average value can be expressed as a function of current 
cross-correlators.
This procedure allows the detection of entanglement, through the measurement of the witness operator, if the latter does not depend on the unknown parameters describing the corruption of the state.
Being a necessary and sufficient criterion for entanglement detection, the method of the witness operator allowed us to prove that spin singlet and triplet electron states are robust against the action of pure dephasing in the system. 
However, entanglement can be destroyed when spin relaxation in the conductors 
is considered.

For the sake of illustration, we applied the optimal witness construction also to
two realistic Hall bar systems: the quantum point contact tunneling barrier and the electronic equivalent of the optical Hanbury-Brown and Twiss interferometer, where 
electron-hole entangled pairs are generated.
Dephasing was introduced through a phenomenological model of randomly distributed phases accumulated by the particles propagating through the edges channels.
Apart from demonstrating the usefulness of the witness method in such setups, we showed that the witness detection can be used as a diagnostic tool in order to infer information about the ``noise'' affecting the system.
Unfortunately, a measurable optimal witness operator could not be found, using the separability criterion, for the tunneling barrier system in the case of arbitrary channel mixing.

Compared to a Bell inequality test, the detection of entanglement by means of optimal witness operators presents two remarkable advantages.
Firstly, being a sufficient and necessary criterion, entanglement is always detected in the considered class of states.
Secondly, independently of the degree of dephasing, the evaluation of the mean value of the witness requires a fixed set of cross-correlated measurements, corresponding to a small number (from 10 to 20) of current cross-correlations.
A drawback is that the method of the witness requires some knowledge of the state produced by a given system (in terms of an expected class of states) and that the procedure for constructing a measurable optimal witness operator is not given a priori.

For the future it would be interesting to find other schemes for constructing optimal witness operators which allow a detection of entanglement even in the cases where the separability criterion fails in producing a measurable witness.
To conclude, we stress that the method of the witness operator presented here can be applied to any systems and we hope that it can be of practical use in experiments.

\appendix

\section{Current operator and current cross-correlators}

\subsection{Entangler prototype setup}
\label{app:pro}
The current operator for electrons flowing through terminal A with spin $\alpha$ along the spin quantization axis identified by the vector $\vec{a}$ is defined as \cite{Blanter00}:
\begin{eqnarray}
I^{\vec{a}}_{\text{A}\alpha}(t)=\frac{e}{h}\int dE dE' e^{i(E-E')t/\hbar} \times\nonumber\\
\left[a^{\dagger}_{\text{A}\vec{a}\alpha}(E) a_{\text{A}\vec{a}\alpha}(E') - b^{\dagger}_{\text{A}\vec{a}\alpha}(E) b_{\text{A}\vec{a}\alpha}(E')\right] ,
\label{curop}
\end{eqnarray}
where $a^{\dagger}_{\text{A}\vec{a}\alpha}(E)$ ($b^{\dagger}_{\text{A}\vec{a}\alpha}(E)$) is the creation operator for incoming (outgoing) electrons with energy $E$. 
At zero temperature, the mean current $\langle I^{\vec{a}}_{\text{A}\alpha} \rangle$ can be calculated as the expectation value of (\ref{curop}) over the outgoing state, whose density matrix is given in Eq.~(\ref{rho}).
As a result,
\begin{equation}
\langle I^{\vec{a}}_{j\alpha}\rangle=\frac{e^2V}{h} \langle N^{\vec{a}}_{j\alpha}\rangle ,
\end{equation}
where ${N^{\vec{a}}_{j\alpha}=b^{\dagger}_{\text{A}\vec{a}\alpha}(E) b_{\text{A}\vec{a}\alpha}(E)}$ is the number operator.

The zero-frequency current cross-correlator, defined in Eq.~(\ref{cross}), can be written in terms of the number operators as \cite{Samuelsson06}:
\begin{eqnarray}
s_{\alpha\beta}^{\vec{a}\vec{b}} (0)= \frac{e^3V}{h}\left [\langle N^{\vec{a}}_{\text{A} \alpha} N^{\vec{b}}_{\text{B} \beta} \rangle 
-   \langle N^{\vec{a}}_{\text{A} \alpha} \rangle \langle N^{\vec{b}}_{\text{B} \beta} \rangle \right ] .
\label{sc}
\end{eqnarray}
For the prototype setup of Sec.~\ref{ideal} we find that:
\begin{equation}
\langle N^{\vec{a}}_{j\alpha}\rangle=\frac{|t_j|^2}{2} .
\end{equation}
At zero temperature, the correlators ${\langle N^{\vec{a}}_{\text{A} \alpha} N^{\vec{b}}_{\text{B} \beta} \rangle}$ can be calculated  as expectation values 
over the outgoing state. Along the z axis, one finds that
\begin{eqnarray}
\langle N^{\vec{z}}_{\text{A} \uparrow (\downarrow)} N^{\vec{z}}_{\text{B} \downarrow (\uparrow)} \rangle&=& \frac{|t_{\text{A}}|^2 |t_{\text{B}}|^2}{2} \; \\
\langle N^{\vec{z}}_{\text{A} \uparrow (\downarrow)} N^{\vec{z}}_{\text{B} 
\uparrow (\downarrow)} \rangle&=&0 ,
\end{eqnarray}
where $\vec{z}$ is the versor of the z axis.
Along the x and y axes one finds that:
\begin{eqnarray}
\langle N^{\vec{a}}_{\text{A} \uparrow (\downarrow)} N^{\vec{a}}_{\text{B} \downarrow (\uparrow)} \rangle&=&\frac{|t_{\text{A}}|^2 |t_{\text{B}}|^2}{4} \left[ 1+(-1)^{s+1}\cos \left (x_{\text{A}}-x_{\text{B}}\right ) \right] \; \notag \\
\langle N^{\vec{a}}_{\text{A} \uparrow (\downarrow)} N^{\vec{a}}_{\text{B} 
\uparrow (\downarrow)} \rangle&=&\frac{|t_{\text{A}}|^2 |t_{\text{B}}|^2}{4} \left[ 1+(-1)^{s} \cos \left (x_{\text{A}}-x_{\text{B}}\right ) \right] \notag
\end{eqnarray}
with $\vec{a}=\vec{x},\vec{y}$.

\subsection{HBT interferometer}
\label{app:hbt}
The zero frequency current cross-correlations between contacts 3 and 4, and 5 and 6 of the HBT can be calculated using the parametrization of the QPC given in Eq.~(\ref{bs}):
\begin{widetext}
\begin{eqnarray}
s_{36}(0)&=&-\frac{2 e^3 V}{h} {\cal R T} \left [ \left (g^A_{11} g^B_{21} \right )^2 + \left (g^A_{21} g^B_{11} \right )^2+ 2 \left (e^{i(\phi_0 + \delta)}-e^{-i(\phi_0 + \delta)} \right ) g^A_{11} g^B_{21} g^A_{21} g^B_{11} \right ]\;\notag\\
s_{45}(0)&=&-\frac{2 e^3 V}{h} {\cal R T} \left [\left (g^A_{12} g^B_{22} \right )^2 + \left (g^A_{22} g^B_{12} \right )^2+ 2 \left (e^{i(\phi_0 + \delta)}-e^{-i(\phi_0 + \delta)} \right ) g^A_{12} g^B_{22} g^A_{22} g^B_{12} \right ]\;\notag\\
s_{35}(0)&=&-\frac{2 e^3 V}{h} {\cal R T} \left [ \left (g^A_{11} g^B_{22} \right )^2 + \left (g^A_{21} g^B_{12} \right )^2- 2  \left (e^{i(\phi_0 + \delta)}-e^{-i(\phi_0 + \delta)} \right ) g^A_{11} g^B_{22} g^A_{21} g^B_{12} \right ]\;\notag\\
s_{46}(0)&=&-\frac{2 e^3 V}{h} {\cal R T} \left [\left (g^A_{12} g^B_{21} \right )^2 + \left (g^A_{12} g^B_{11} \right )^2- 2  \left (e^{i(\phi_0 + \delta)}-e^{-i(\phi_0 + \delta)} \right ) g^A_{12} g^B_{21} g^A_{12} g^B_{11} \right ]\;
\label{zcorbut}  
\end{eqnarray}
\end{widetext}
where $g^{\text{A(B)}}_{ij}$ denotes the element $(i,j)$ of the matrix $G_{\text{A(B)}}$ of Eq.~(\ref{bs}).
Notice, in particular, that such current correlations depend on $\theta_{i}$ and the phases $\varphi^i_{jk}$.

We define the zero frequency cross-correlators ${s^{\vec{a}\vec{b}}_{\alpha\beta}}(0)$,
where ${\vec{a},\vec{b}=\vec{x},\vec{y},\vec{z}}$, $\alpha=3,4$ and $\beta=5,6$, 
by choosing different configurations for the QPCs $\text{A}$ and $\text{B}$.
Indeed, by setting
\begin{eqnarray}
&1.&~\theta_i=0;~\varphi_{11}=0;~\varphi^i_{22}=\pi;~\varphi^i_{12}=\varphi^i_{21}=0\notag 
\\
&2.&~\theta_i=\frac{\pi}{4};~\varphi^i_{11}=0;~\varphi^i_{22}=\pi;~\varphi^i_{12}=\varphi^i_{21}=0 \\
&3.&~\theta_i=\frac{\pi}{4};~\varphi^i_{22}=\frac{\pi}{2};~\varphi^i_{21}=-\frac{\pi}{2};~\varphi^i_{11}=\varphi^i_{12}=0
\notag
\end{eqnarray}
we obtain : 
\begin{eqnarray}
&1.&~G_i = Z \equiv \left ( \begin{array}{cc}
1& 0 \\
0 & -1
\end{array} \right )\;\notag\\
&2.&~G_i=X \equiv\frac{1}{\sqrt{2}}\left ( \begin{array}{cc}
1& 1 \\
1 & -1
\end{array} \right )\label{1}\\
&3.&~G_i=Y \equiv\frac{1}{\sqrt{2}}\left ( \begin{array}{cc}
1& 1 \\
-i & i
\end{array} \right )\; \notag 
\end{eqnarray}
corresponding, respectively, to the settings $\vec{z}$, $\vec{x}$ and $\vec{y}$.
For example, we find
\begin{eqnarray}
s_{36}^{\vec{x}\vec{y}}(0)&=&-\frac{2 e^3 V}{h} {\cal R T} \left [\left (X_{11} Y_{21} \right )^2 + \left (X_{21} Y_{11} \right )^2 \right . \;\notag \\
&+& \left .  2\left (e^{i(\phi_0 + \delta)}-e^{-i(\phi_0 + \delta)} \right ) X_{11} Y_{21} X_{21} Y_{11}\right ] \;\notag\\
&=&-\frac{e^3 V}{h} {\cal R T} \left [1+ \sin \left (\phi_0+\delta \right ) \right ] ,
\end{eqnarray}
$X_{ij}$ ($Y_{ij}$) being the elements of the matrix $X$ ($Y$) of Eq.~(\ref{1}).

\begin{acknowledgments}
We thank C. Macchiavello and especially Rosario Fazio 
(who participated in this work at the early stages) for useful discussions. 
This work was supported by EC through grants EC-RTN Nano, 
EC-RTN Spintronics and EC-IST-SQUIBIT2 (F.T.) 
and by the National Security Agency (NSA) under Army Research Office 
(ARO) contract number W911NF-06-1-0208 (L.F.).
\end{acknowledgments}

\bibliography{eqip}

\begin{thebibliography}{54}
\bibitem[{not({\natexlab{a}})}]{nota1}
\bibinfo{note}{It is interesting to calculate the statistical fluctuations of
  $W$, whose root mean square is defined as ${\langle \delta
  W\rangle=\sqrt{\langle W^2\rangle-\langle W\rangle^2}}$. A simple calculation
  yields ${\langle \delta W\rangle=\frac{1}{2}\sqrt{1-e^{-4\omega^2}}}$. Note
  that the absolute mean value of $W$ equals $\langle \delta W\rangle$ at
  $\omega=1/2\sqrt{\log 2}$.}
\bibitem[{not({\natexlab{b}})}]{nota3}
\bibinfo{note}{A finite measurement time implies an energy uncertainty given by
  $h/t_{\text{m}}$, so that the number of energy intervals of distinguishable
  energy is $eVt_{\text{m}}/h$.}
\bibitem[{not({\natexlab{c}})}]{nota2}
\bibinfo{note}{Note that the mean number of particles detected over a measuring
  time $t_{\text{m}}$ in A is given by ${\langle
  Q_{\text{A}\vec{a}\alpha}(t_{\text{m}}) \rangle=\int_0^{t_{\text{m}}} dt
  \langle I_{\text{A}\alpha}^{\vec{a}}(t) \rangle}$}.
\bibitem[{foo({\natexlab{b}})}]{footnoteNo2}
\bibinfo{note}{Note that, in our setup,
  $\sum_{\alpha\beta}s_{\alpha\beta}^{\vec{a}\vec{b}}(0)=0$.}
\end{thebibliography}

\end{document}